\documentclass[12pt]{iopart}

\usepackage{cite}
\usepackage[utf8]{inputenc}
\usepackage{graphicx}
\usepackage{graphicx,epstopdf}
\usepackage{dcolumn}
\usepackage{amsmath}
\usepackage{lineno,hyperref}
\usepackage{amsfonts}
\usepackage{amsthm}
\usepackage{bbm}
\usepackage{amssymb}
\usepackage{mathrsfs}
\PassOptionsToPackage{caption=false}{subfig}
\usepackage{subfig}
\usepackage{color}
\usepackage[normalem]{ulem}

\newcommand{\1}{\mathbbm{1}}

\newcommand{\Z}{\mathbbm{Z}}

\newcommand{\Ocal}{\mathcal{O}}

\newcommand{\ket}{\rangle}
\newcommand{\bra}{\langle}

\begin{document}

\title[A. C. Santos, Quantum gates by inverse engineering of a Hamiltonian]{Quantum gates by inverse engineering of a Hamiltonian}

\author{Alan C. Santos}

\address{Instituto de F\'isica, Universidade Federal Fluminense \\ Av. Gal. Milton Tavares de Souza s/n, Gragoat\'a, 24210-346, Niter\'oi, RJ, Brazil}
\ead{ac\_santos@id.uff.br}
\vspace{10pt}
\begin{indented}
\item[]October 2017
\end{indented}

\begin{abstract}
Inverse engineering of Hamiltonian (IEH) from an evolution operator is a useful technique for protocol of quantum control with potential applications in quantum information processing. In this paper we introduce a particular protocol to perform IEH and we show how this scheme can be used for implementing a set of quantum gates by using minimal quantum resources (such as entanglement, interactions between more than two quits or auxiliary quits). Remarkably, while previous protocols request three-quits interactions and/or auxiliary quits for implementing such gates, our protocol requires just two-qubit interactions and no auxiliary qubits. By using this approach, we can obtain a large class of Hamiltonians that allow us to implement single and two-quit gates necessary to quantum computation. To conclude this article, we analyze the performance of our scheme against systematic errors related to amplitude noise, where we show that the free parameters introduced in our scheme can be useful for enhancing the robustness of the protocol against such errors.
\end{abstract}

\section{Introduction}

Currently, protocols of quantum control with time-dependent Hamiltonians like adiabatic passage \cite{Gaubatz:90}, Lewis-Riesenfeld invariants \cite{Lewis:67,Lewis:69}, transitionless quantum driving (TQD) \cite{Demirplak:03,Demirplak:05,Berry:09} and the current proposal of the inverse engineering of a Hamiltonin (IEH) from unitary evolutions operators \cite{Kang:16}, have played an important role in quantum information processing (see \cite{Torrontegui:13} for a detailed review of many applications of the three first techniques). In addition, in the few last years, many experimental and theoretical studies have been performed in order to analyze the robustness of such protocols against decoherence effects \cite{Childs:01,AminPRA:09,Du:16,Adolfo:16,Santos:17,Chen-Huang:16,XiaPRA:14,Chen-Huang:15,Lu-Xia:14}.

Quantum control via time-dependent Hamiltonians is of great interest for many knowledge fields in physics, in particular for quantum computation (QC) and information (our focus in this paper). For example, these techniques are used for solve problems of satisfiability via adiabatic dynamics \cite{Farhi:01}, engineering fast Hamiltonians for speeding up QC \cite{SciRep:15,PRA:16,Front:16} and state preparation in quantum simulations of relativistic dynamics \cite{Song:17}, for example. In addition, we can use such protocols for developing hybrid schemes of quantum computation, where we obtain controllable time-dependent Hamiltonians to implement single and controlled quantum gates. For example, quantum gates of a circuit can be implemented via adiabatic Hamiltonians \cite{Bacon:09,Hen:15}, via counter-diabatic dynamics \cite{SciRep:15,PRA:16,Front:16} and via shortcuts to adiabatic holonomic QC with transitionless quantum driving dynamics \cite{Song:16}. However, in order to implement \textit{universal} QC, these schemes requires auxiliary qubits, many body interactions, etc \cite{Bacon:09,Hen:15,SciRep:15,PRA:16,Front:16}. We say \textit{universal} in the sense that, given an \textit{unknown} input state, we should be able to implement any single- and two-qubit quantum gate on such a qubit.

In this paper we introduce an alternative way of obtaining Hamiltonians for implementing quantum gates based on IEH from evolution operators. Different from methods previously developed for IEH and shortcuts to adiabaticity, ancilla qubits or highly degenerate Hamiltonians are not necessary for our scheme. In the Sec. \ref{discuss} we discuss the general aspects of our approach and we show how to obtain a set of Hamiltonian which allow us to implement the quantum gates of a quantum circuit. In this sense, the scheme present here is an enhanced way to implement quantum gates without auxiliary resources. In the Sec. \ref{application} we illustrate the results obtained here by providing a set of Hamiltonian for implementing a restricted set of quantum gates necessary for QC \cite{Nielsen:Book,Hen:14}.

\section{Inverse Engineering of a Hamiltonian} \label{discuss}

Let us start by considering the Schrödinger equation (we set $\hbar=1$ throughout the manuscript)
\begin{eqnarray}
H(t)|\psi(t)\ket = i |\dot{\psi}(t)\ket \mathrm{ . }
\end{eqnarray}
For a unitary dynamics, there is an operator $U(t)$ that allow us to write $|\psi(t)\ket = U(t)|\psi(0)\ket$. From equation above (valid for any $|\psi(0)\ket$), the Hamiltonian reads as
\begin{eqnarray}
H(t) = i \dot{U}(t) U^{\dagger}(t) \mathrm{ , } \label{Hstart}
\end{eqnarray}
the well-known equation to obtain the Hamiltonian associated with the evolution operator $U(t)$ \cite{Messiah:Book,Nielsen:Book}. This equation is the starting point for protocols of inverse engineering in closed quantum systems \cite{Marcela:14,Jing:13,Kang:16}, as well as transitionless quantum driving \cite{Demirplak:03,Demirplak:05,Berry:09,Torrontegui:13}. In alternative approaches, the operator $U(t)$ has been considered as
\begin{eqnarray}
U^{\prime}(t) = | k(t) \ket \bra k(0)| + \sum \nolimits _{m,n \neq k} \lambda_{mn}(t) | m(t) \ket \bra n(0)| \mathrm{ , }
\end{eqnarray}
where $| n(t) \ket$ is a complete orthonormal basis for the Hilbert space of the system and $\lambda_{mn}(t)$ are free parameters. Therefore, any system driven by the Hamiltoinian $H^{\prime}(t) = i \dot{U^{\prime}}(t) U^{\prime\dagger}(t)$ is begun in the state $| k(0) \ket $ and evolves to $| k(\tau) \ket $ through path $| k(t) \ket $, with $0 \leq t \leq \tau$. Since we can obtain the transitionless theory from a suitable choice of the parameters $\lambda_{mn}(t)$, we can consider the operator $U^{\prime}(t)$ as the most general form of $U(t)$ \cite{Kang:16}.

On the other hand, our alternative approach of Hamiltonian engineering is obtained from a new definition of the operator $U(t)$ as
\begin{eqnarray}
U(t) = \sum \nolimits _{n} e^{i \varphi_{n}(t)}| n(t) \ket \bra n(t)| \mathrm{ , }
\end{eqnarray}
where $| n(t) \ket$ constitutes an orthonormal bases for the Hilbert space associated with the system and $\varphi_{n}(t)$ are real free parameters. It is easy to show that $U(t)$ satisfies the unitarity condition $U(t) U^{\dagger}(t) = \1$ for any set of parameters $\varphi_{n}(t)$. In addition, for obtaining an operator that satisfies the initial condition $U(0) = \1$, we must impose initial conditions for the parameters $\varphi_{n}(t)$ given by $\varphi_{n}(0)=2n\pi$ for $n \in \Z$.

Differently from others protocols \cite{Gaubatz:90,Lewis:67,Lewis:69,Demirplak:03,Demirplak:05,Berry:09,Chen-Huang:16,Chen-Huang:15,Chen:SREP16,Kang:16}, we can see that our definition of the operator $U(t)$ is an operator most general than some operator that drives the system from a known initial state $| n(0) \ket$ to $| n(\tau) \ket$. Therefore, this method is not dependent on the initial state $| n(0) \ket$. As we shall see, this approach can be useful in some protocols of quantum information processing, e.g. to implement a set of quantum gates necessary for universal QC using minimal resource.

\section{Quantum gates by Inverse Engineering of a Hamiltonian}

\subsection{Single-qubit gates}

In this section we will show how single quantum gates can be implemented, without additional resources, by using the scheme presented here. To this end, let us consider that a single-qubit gate can be view as a linear transformation on an arbitrary quantum state $| \psi _{\mathrm{inp}} \ket = a|0\ket+b|1 \ket$, and so let us consider the transformation $| \psi (t) \ket = U_{1}(t)| \psi _{\mathrm{inp}} \ket$, where the operator $U_{1}(t)$ is given by
\begin{eqnarray}
U_{1}(t) = | n_{+}(t) \ket \bra n_{+}(t) | + e^{i\varphi (t)}| n_{-}(t) \ket \bra n_{-}(t) | \mathrm{ , } \label{U1}
\end{eqnarray}
where
\begin{eqnarray}
| n_{+}(t) \ket &=& \cos [\theta (t)/2] | 0 \ket + e^{i\phi (t)} \sin [\theta (t)/2] | 1 \ket \mathrm{ , } \\
| n_{-}(t) \ket &=& e^{i\phi (t)} \cos [\theta (t)/2] | 1 \ket - \sin [\theta (t)/2] | 0 \ket \mathrm{ , }
\end{eqnarray}
with $\theta(t)$, $\varphi (t)$ and $\phi (t)$ being real free parameters. It is easy to show that the conditions $U_{1}(t) U_{1}^{\dagger}(t) = \1$ and $U_{1}(0) = \1$ are satisfied if we choose $\varphi (t)$ such that $\varphi (0)= 2 n \pi$, for $n$ integer. Parameters associated with the quantum gate to be implemented are encoded in the parameters $\theta(t)$, $\varphi (t)$ and $\phi (t)$. To show that we can really implement single-qubit gates by using the operator $U_{1}(t)$, let us consider an arbitrary input state $| \psi _{\mathrm{inp}} \ket$ so that the evolved state $|\psi(t)\ket$ is given by
\begin{eqnarray}
| \psi (t) \ket = U_{1}(t)| \psi _{\mathrm{inp}} \ket = \alpha (t) |0 \ket + \beta (t) |1 \ket \mathrm{ , } \label{psi1Evol}
\end{eqnarray}
where the coefficients $\alpha (t)$ and $\beta (t)$ are given, respectively by
\begin{eqnarray}
\alpha (t) &=& \frac{a \sigma_{+}(t) - \sigma_{-}(t)\tilde{\alpha}(t)}{2} \mathrm{ \ \ \ , \ \ \ } \label{alpha1-beta1}
\beta (t) = \frac{ b \sigma_{+}(t) + \sigma_{-}(t)\tilde{\beta}(t)}{2} \mathrm{ , }
\end{eqnarray}
with $\sigma_{\pm}(t)=(e^{i\varphi (t)} \pm 1)$, $\tilde{\alpha}(t) = a\cos \theta(t) + b e^{- i\phi (t)} \sin \theta(t)$ and $\tilde{\beta}(t) =b\cos \theta(t) - a e^{i\phi (t)} \sin \theta(t)$. By using the initial condition $\varphi (0)= 2 n \pi$ we can see that $\alpha (0)=a$ and $\beta (0) = b$, because $\sigma_{\pm}(0)=2\delta_{+\pm}$. Therefore, from Eqs. (\ref{psi1Evol}-\ref{alpha1-beta1}), an arbitrary single-qubit rotation can be performed. 

Notice that we have implemented an arbitrary rotation on an unknown input state $|\psi _{\mathrm{inp}} \ket$. Thus, arbitrary single-qubit universal operations can be performed by using this approach, where no additional quantum resource (such as entanglement or auxiliary qubits, for example) is required. In addition, as we shall see  in the next section, this model allow us find  both trivial and nontrivial Hamiltonians to implement a same gate.

\subsection{Two-qubit quantum gates}

To show that our protocol can be used to implement an universal set of quantum gates for quantum computation, we must show how to implement two-qubit quantum gates. To this end, let us write the two-qubit input state generically as $|\psi _{\mathrm{inp},2} \ket = a|00\ket+b|01\ket+c|10\ket+d|11\ket$. In addition, we define the operator
\begin{eqnarray}
U_{2}(t) = \sum _{k=1,2} | n_{k,+}(t) \ket \bra n_{k,+}(t) | + e^{i\varphi_{k} (t)}| n_{k,-}(t) \ket \bra n _{k,-} (t) \vert \mathrm{ , } \label{U2}
\end{eqnarray}
where (with $\bar{k} =k-1$)
\begin{eqnarray}
| n_{k,+}(t) \ket &=& \cos [\theta _{k}(t)/2] | \bar{k}0 \ket + e^{i\phi _{k} (t)} \sin [\theta _{k}(t)/2] | \bar{k}1 \ket \mathrm{ , } \label{nkmais} \\ 
| n_{k,-}(t) \ket &=& e^{i\phi _{k} (t)} \cos [\theta _{k}(t)/2] | \bar{k}1 \ket - \sin [\theta _{k}(t)/2] | \bar{k}0 \ket \label{nkmenos} \mathrm{ , }
\end{eqnarray}
with the initial conditions $\varphi _{1} (0)=\varphi _{2} (0)=2 n \pi$ (due to requirement $U_{2}(0) = \1$). Now we have six free parameters, so that we will use them in order to obtain arbitrary two-qubit operation. It is easy to see that the operator $U_{2}(t)$ is a general two-qubit operator. Therefore we can adjust adequately our free parameters for obtaining $U_{2}(t)$ as an entangled gate or a composition of two independent single-qubits gates  (i.e., $U_{2}(t) = A_{1}(t) \otimes A_{2}(t)$). In order to analyze some results by using its most general form, we keep our discussion without consider some particular case for $U_{2}(t)$, but particularizations for $U_{2}(t)$ will be taking into account in the Sec. \ref{application}.

From Eqs. (\ref{U2}-\ref{nkmenos}) we can write the evolved state as
\begin{eqnarray}
|\psi _{\textmd{{\scriptsize inp}},2} \ket &=& U_{2}(t) |\psi _{\mathrm{inp},2} \ket = \alpha (t) |00\ket + \beta (t) |10\ket + \gamma (t) |01\ket + \delta (t) |11\ket \mathrm{ , }
\end{eqnarray}
with the following coefficients
\begin{eqnarray}
\alpha (t) &=& \frac{a \sigma_{1,+}(t) - \sigma_{1,-}(t)\tilde{\alpha}(t)}{2} \mathrm{ \ \ \ , \ \ \ } \beta (t) = \frac{b \sigma_{1,+}(t) + \sigma_{1,-}(t)\tilde{\beta}(t)}{2} \label{alpha2-beta2} \mathrm{ , } \\
\gamma (t) &=& \frac{c \sigma_{2,+}(t) - \sigma_{2,-}(t)\tilde{\gamma}(t)}{2} \mathrm{ \ \ \ , \ \ \ } \delta (t) = \frac{d \sigma_{2,+}(t) + \sigma_{2,-}(t)\tilde{\delta}(t)}{2} \label{gama-delta} \mathrm{ . }
\end{eqnarray}
Again we have defined $\sigma_{k,\pm}(t)=(e^{i\varphi _{k} (t)} \pm 1)$, $\tilde{\alpha}(t) = a\cos \theta _{1}(t) + b e^{i\phi _{1}(t)} \sin \theta _{1}(t)$, $\tilde{\beta}(t) = b\cos \theta _{1}(t) - a e^{-i\phi _{1}(t)} \sin \theta _{1}(t)$, $\tilde{\gamma}(t)=c\cos \theta _{2}(t) + d e^{i\phi _{2}(t)} \sin \theta _{2}(t)$ and $\tilde{\delta}(t)=d\cos \theta _{2}(t) - c e^{-i\phi _{2}(t)} \sin \theta _{2}(t)$. Thus, an arbitrary two-qubit gate can be implemented with this scheme.

In this discussion we have not labeled the target and control qubit, however this choice can be done through the definition of the free parameters. The operator $U_{2}(t)$ encompasses a large class of two-qubit gates, i.e., $U_{2}(t)$ can be an entangling quantum gates (as CNOT or some controlled single-qubit unitary rotations) or non-entangling gates (such as the SWAP gate).

\section{Quantum gates for (approximately) universal QC} \label{application}

In order to show how we can implement a set of universal quantum gates by using the results developed here, in this section we consider some choices for the free parameters previously discussed. As application of this method, we discuss about Hamiltonians able to implement a set of quantum gates necessary for implementing universal QC with arbitrary precision, namely, the set $\{H, S, T, CZ\}$, where $H$ represents the Hadamard gate, $S$ and $T$ are $\pi/4$ and $\pi/8$ gates, respectively, and $CZ$ is the controlled-phase gate \cite{Nielsen:Book,Barenco:95}.

\subsection{Single-qubit gates} \label{SecSingleGates}

The Hamiltonian for implement single quantum gates, that can be obtained from Eq. (\ref{Hstart}), is \textit{not} trivial and written as (in order to obtain simple Hamiltonians, throughout the manuscript we consider that the system evolves up to a global phase)
\begin{eqnarray}
H(t) = \frac{1}{2} \vec{\omega}(t) \cdot \vec{\sigma} \mathrm{ , } \label{GenericH}
\end{eqnarray}
where $\vec{\sigma}$ is a ``vector" with its components given by the Pauli matrices $\sigma _{x}$, $\sigma _{y}$ and $\sigma _{z}$, and $\vec{\omega}(t)$ is a vector where its components are given by
\begin{eqnarray}
\omega_{x}(t) &=& ( \cos \varphi -1 ) \dot{ \phi } \cos \phi \cos \theta \sin \theta 
+ ( \dot{\theta} \cos \theta \sin \varphi + \dot{\varphi} \sin \theta )\cos \phi
\nonumber \\
&+& [  \dot{\phi} \sin \theta \sin \varphi + ( \cos \varphi -1 ) \dot{\theta} ] \sin \phi  \mathrm{ , } \label{omegaX} \\
\omega_{y}(t) &=& ( \cos \varphi -1 ) \dot{ \phi } \sin \phi \sin \theta \cos \theta 
+ \sin \phi ( \dot{\theta} \cos \theta \sin \varphi + \dot{\varphi} \sin \theta ) 
\nonumber \\
&+& [ \dot{\phi} \sin \theta \sin \varphi - ( \cos \varphi -1 ) \dot{\theta} ] \cos \phi \mathrm{ , } \label{omegaY} \\
\omega_{z}(t) &=& - \dot{\theta} \sin \theta \sin \varphi  -( \cos \varphi -1 ) \dot{\phi} \sin ^{2}\theta  + \dot{\varphi} \cos \theta \label{omegaZ} \mathrm{ . }
\end{eqnarray}

Therefore, now we are enable to particularize the Hamiltonian of the Eq. (\ref{GenericH}) in order to obtain a restricted set of Hamiltonian associated with quantum gates used for implement universal QC (approximately) \cite{Nielsen:Book,Barenco:95}.

\vspace{0.3cm}

\textbf{Hamiltonian for phase shift gates --} For phase shift gates, given any entry $|\psi_{\mathrm{inp}} \ket = a | 0 \ket + b | 1 \ket$ we have the corresponding output $|\psi_{out} \ket = a | 0 \ket + e^{i \xi}b | 1 \ket$, for an arbitrary value $0 < \xi < 2 \pi$. Thus, from Eq. (\ref{alpha1-beta1}), we see that such gate is implemented if we choose $\varphi (\tau) = \xi$ and $\theta (\tau) = 2n\pi$, with $n \in \Z$ and $\tau$ being the total evolution time. Since we have boundary conditions for the parameter $\varphi (t)$, namely $\varphi (0) = 0$ and $\varphi (\tau) = \xi$, so $\varphi (t)$ can not assume an arbitrary form. On the other hand, we see that no consideration has been done about the parameter $\phi(t)$, so that we can consider it arbitrary. For simplicity we will consider that $\phi(t)=0$. Therefore, the components of $\vec{\omega}^{\mathrm{ph}}(t)$ associated with the Hamiltonian $H_{\mathrm{ph}}(t) = (1/2) \vec{\omega}^{\mathrm{ph}}(t) \cdot \vec{\sigma}$ becomes
\begin{eqnarray}
\omega_{x}^{\mathrm{ph}}(t) &=&  \cos (\theta ^{\mathrm{ph}}) \sin (\varphi ^{\mathrm{ph}}) \dot{\theta}^{\mathrm{ph}} + \sin (\theta^{\mathrm{ph}}) \dot{\varphi} ^{\mathrm{ph}} \mathrm{ , } \label{omegaXPhaseGate} \\
\omega_{y}^{\mathrm{ph}}(t) &=& - [ \cos (\varphi^{\mathrm{ph}}) -1 ] \dot{\theta}^{\mathrm{ph}} \mathrm{ , } \label{omegaYPhaseGate}\\
\omega_{z}^{\mathrm{ph}}(t) &=& \cos (\theta^{\mathrm{ph}}) \dot{\varphi}^{\mathrm{ph}} - \sin (\theta^{\mathrm{ph}}) \sin (\varphi^{\mathrm{ph}}) \dot{\theta} \label{omegaZPhaseGate} \mathrm{ , }
\end{eqnarray}
where we have labeled the parameters by using ``ph" in order to explicit that these parameters depends on the gate to be implemented.

For some experimental architectures the operation $\sigma_{y}$ is not easily implementable, for example in systems composed by Bose–Einstein condensates in optical lattices \cite{Bason:12}, experimental architecture of superconducting circuits \cite{Johnson:11,Harris:10,Orlando:99,You:05}. Thus, such experimental difficulty is not a problem if we set $\theta^{\mathrm{ph}} (t) = \theta_{0} = \mathrm{cte}$. 

Therefore, by taking into account those considerations related with parameters $\phi^{\mathrm{ph}}(t)$, $\theta^{\mathrm{ph}}(t)$ and $\varphi(t)$, let us put $\theta^{\mathrm{ph}} (t) = \phi^{\mathrm{ph}}(t) = 0$, so that the Hamiltonian for phase shift gates is given by 
\begin{eqnarray}
H_{\mathrm{ph}}(t) = \frac{\dot{\varphi}^{\mathrm{ph}}(t)}{2} \sigma_{z} \mathrm{ . }
\end{eqnarray}

The Hamiltonian above can be implemented in nuclear magnetic resonance (NMR) experimental setups, where a magnetic $\vec{B}$ field is used for driving nuclear spins of atoms and molecules. In general such field is taken constant \cite{Sarthour:Book}, thus we can set $\varphi(t) = \xi t/ \tau$, where $\xi$ is the shift phase to be implemented and $\tau$ is the total evolution time. Therefore, we obtain the time-independent Hamiltonian
\begin{eqnarray}
H_{\mathrm{ph}}(t) = \frac{\xi}{2\tau} \sigma_{z} \mathrm{ , } 
\end{eqnarray}
where $1/\tau$ can be identified as the Larmor frequency $\omega_{0} \propto \gamma_{0} B_{z} $ of a nuclear spin, where the magnetic field is $\vec{B} = B_{z} \hat{z}$ and $\gamma_{0}$ is the gyromagnetic ratio of the nucleus. In order to give a Thus, we can set the total evolution time from intensity of the magnetic field $\vec{B}$. In particular, if we put $\xi = \xi_{\mathrm{S}} = \pi/2$, $\xi = \xi_{\mathrm{T}} = \pi/4$ and $\xi = \xi_{\mathrm{Z}} = \pi$, we obtain the Hamiltonian that implements the $S$, $T$ and $Z$ gates \cite{Nielsen:Book}, respectively.

It would be worth mention that this choice choice is not unique and we can have many others possibilities if we set $\theta^{\mathrm{ph}} (t) \neq 0$. However, if we pick $\theta^{\mathrm{ph}} (t) \neq 0$ the corresponding Hamiltonian will be not as simple as the Hamiltonian obtained above. In conclusion, we have showed that our approach allow us to find both trivial and nontrivial Hamiltonians to perform a same task. Moreover, we can obtain time-independent Hamiltonians feasible in the lab.

\vspace{0.3cm}

\textbf{Hamiltonian for Hadamard gate --} The Hadamard gate is an exclusive gate of quantum computers due its particular task of generating quantum superpositions with elements of the computational basis. More specifically, given a quantum state $|\psi_{\mathrm{inp}} \ket = a | 0 \ket + b | 1 \ket$, we get its corresponding output $|\psi_{out} \ket = (a+b) / \sqrt{2} | 0 \ket + (a-b) / \sqrt{2} | 1 \ket$. To implement such operation we set $\theta (\tau) = \pi /4$, $\phi (\tau) = 0$ and $\varphi (\tau) = \pi$. In this case, there are not free parameters, but due to the boundary condition on $\phi (t)$, we can consider $\phi (t) = 0$ in order to simplify the Hamiltonian $H_{\mathrm{Had}}(t)=\frac{1}{2} \vec{\omega}^{\mathrm{Had}}(t) \cdot \vec{\sigma}$ that implement a Hadamard gate. Under this choice, from Eqs. (\ref{omegaX}), (\ref{omegaY}) and (\ref{omegaZ}), we can see that $\vec{\omega}^{\mathrm{Had}}(t)=\vec{\omega}^{\mathrm{ph}}(t)$, but with different boundary conditions for the parameters $\theta$ and $\varphi$. Thus, we have the set $\{ \omega_{x}^{\mathrm{Had}}(t),\omega_{y}^{\mathrm{Had}}(t),\omega_{z}^{\mathrm{Had}}(t)\}$ given by Eqs. (\ref{omegaXPhaseGate}-\ref{omegaZPhaseGate}), where now the functions $\theta^{\mathrm{ph}}$ and $\varphi^{\mathrm{ph}}$ must satisfy $\theta (\tau) = \pi /4$ and $\varphi (\tau) = \pi$, respectively. In particular, if we set $\theta (t) = \pi /4$, we find
\begin{eqnarray}
H_{\mathrm{Had}}(t) = \frac{\dot{\varphi}(t)}{2 \sqrt{2}} (\sigma_{z} + \sigma_{x}) \mathrm{ , } \label{HadamardHamiltonian}
\end{eqnarray}
where $\varphi(t)$ is an arbitrary function that satisfies the conditions $\phi (\tau) = 0$ and $\varphi (\tau) = \pi$. For example, we can pick $\varphi(t) = \pi t/\tau$. To describe how we can implement this Hamiltonian, let us consider a NMR experimental setup, where we have a time-dependent magnetic field $\vec{B}\left( t\right) =B_{z}\hat{\imath}+B_{\mathrm{RF}}(t)$, where $B_{\mathrm{RF}}(t)$ is a transverse (rotating) field, called radio-frequency field, given by $B_{\mathrm{RF}}(t)= B_{1}\left[ \cos \left( \omega
t\right) \hat{\jmath}+\sin \left( \omega t\right) \hat{k}\right]$, where $\omega$ is the frequency of such field. It is possible to show that the Hamiltonian Eq. (\ref{HadamardHamiltonian}) can be implemented/simulated using the magnetic field $\vec{B}\left( t\right)$, once we set the frequency $\omega$ of the radio-frequency field $B_{1}$ near to the resonance, i.e, if we put $\omega \approx \omega_{0}$. In fact, in the rotating frame the Hamiltonian of the system can be written as a Landau-Zener Hamiltonian. The demonstration of such result can be found from the Ref. \cite{Nielsen:Book}, see section \textit{7.7 -- Nuclear magnetic resonance}, page 326.

\subsection{Controlled phase shift gate}

A phase controlled phase shift gate is a two-qubit gate that introduces a phase $e^{i \xi}$ controlled by one qubit, where for any input state given by $a|00\ket +b|01\ket +c|10\ket +d|11\ket $ the output state is $a|00\ket +b|01\ket +c|10\ket +e^{i \xi}d|11\ket $. In particular, for $\xi = \pi$ we get the CZ gate (controlled-phase gate, also known as CPHASE or CSign). In general, the CZ is an gate that naturally can be implemented Linear Optical Quantum Computing (LOQC) \cite{Pieter:07} and constitutes a required gate for universal quantum computing \cite{Nielsen:Book}. In addition, we can use CZ gate and single qubit gates to implement a CNOT gate in different experimental architectures \cite{Chatterjee:15,Nielsen:Book,Filidou:12}. Here we will consider the simplest Hamiltonian for implementing such gate. 

Without loss of generality, let us consider a bipartite system initially in the state
\begin{eqnarray}
|\psi _{\mathrm{\mathrm{inp},2}} \ket = a|0 \ket _{\mathrm{c}} | 0\ket _{\mathrm{t}} + b|0 \ket _{\mathrm{c}} | 1\ket _{\mathrm{t}} + c|1 \ket  _{\mathrm{c}} | 0\ket _{\mathrm{t}}+ d|1 \ket _{\mathrm{c}} | 1\ket _{\mathrm{t}} \mathrm{ , } \label{PsiInp2}
\end{eqnarray}
where the subscript ``c" and ``t" labels the control and target qubit, respectively. Thus, under this encoding the state of the system at the end of the evolution can be written as
\begin{eqnarray}
|\psi _{\mathrm{out,2}} \ket = a|0 \ket _{\mathrm{c}} | 0\ket _{\mathrm{t}} + b|0 \ket _{\mathrm{c}} | 1\ket _{\mathrm{t}} + \gamma (\tau)|1 \ket  _{\mathrm{c}} | 0\ket _{\mathrm{t}}+ \delta (\tau)|1 \ket _{\mathrm{c}} | 1\ket _{\mathrm{t}} \mathrm{ , } \label{CNOT}
\end{eqnarray}
where $\gamma (t)$ and $\delta (t)$ are given by the Eq. (\ref{gama-delta}), respectively. Now, we can discuss about the parameters $\varphi_{k}(t)$, $\theta_{k}(t)$ and $\phi_{k}(t)$ necessary for obtain such gate. Firstly, because the coefficients $a$ and $b$ were not changed, from Eq. (\ref{alpha2-beta2}) we conclude such evolution can be achieved of we set the parameter $\varphi_{1}(t)=0$, where no condition about $\theta_{1}(t)$ and $\phi_{1}(t)$ is necessary, thus $\theta_{1}(t)$ and $\phi_{1}(t)$ becomes additional free parameters that can be used to simplify the Hamiltonian. Secondly, to obtain a correct CZ operation, we need to choose our functions so that $\gamma (\tau) = c$ and $\delta (\tau)=e^{i\xi}d$. From Eq. (\ref{gama-delta}), this result can be achieved if we choose $\theta_{2}(\tau) = 0$ and $\varphi_{2}(\tau) = \xi$. Therefore, we have two free parameters that can be used for obtain feasible Hamiltonians, namely, $\phi_{2}(t)$ and $\theta_{2} (t)$.

In particular we can obtain a familiar Hamiltonian if we pick $\phi_{2}(t) = \theta_{2} (t)= 0$, where the corresponding Hamiltonian is written as
\begin{eqnarray}
H(t) = \frac{\dot{\varphi} _{2}(t)}{4} \left[ \1_{\mathrm{c}} \otimes \sigma_{z\mathrm{t}} + \sigma_{z\mathrm{c}} \otimes \1_{\mathrm{t}} - \sigma_{z\mathrm{c}} \otimes \sigma_{z\mathrm{t}} \right] \mathrm{ , } \label{HamCZ}
\end{eqnarray}
for an arbitrary function $\varphi(t)$ satisfying the boundary conditions $\varphi_{2}(0) = 0$ and $\varphi_{2}(\tau) = \xi$.
Remarkably, we can see that the Hamiltonian above requires an interaction $ZZ$ between the physical qubits of the system. Such interaction is a common interaction between nuclear spins present in NMR experimental setups \cite{Chuang:05,Bellac:Book}, therefore the Hamiltonian in Eq. (\ref{HamCZ}) can be implemented for such physical systems.

\section{Robustness against systematic errors}

Now, we will explore the free parameters in order to show how such parameters can be useful in our model for providing robustness against systematic errors. To this end, we will study the stability of our protocol against deviations of physical parameters of the Hamiltonian. Basically, here we will follow the general formalism for such errors in two-level systems, as detailed in Ref. \cite{Ruschhaupt:12}, where the authors have studied protocols where we can find good parameters in order to cancel systematic errors. Unlike from \cite{Ruschhaupt:12}, our protocol allow us to find parameters in order to minimize such systematic errors. In particular, we will consider systematic errors associated with Rabi frequency, that can be simply described by the Hamiltonian $H_{\mathrm{se}}(t) = \omega_{x}(t) \sigma_{x}/2$, where $\omega_{x}(t)$ is the $x$-component of the ideal Hamiltonian $H(t)$ given by Eq. (\ref{GenericH}). In general, such systematic errors are related with deviations in the amplitude of the field from an ideal value. These errors are very common in Hamiltonians driven by laser fields \cite{Ivanov:15,Low:14} and nuclear magnetic resonance \cite{Mitra:09,Raitz:15,Bernardes:16,Isabela:16}, for example. Therefore, the dynamics is given by
\begin{equation}
|\dot{\psi} (t) \ket = [H(t) + \varepsilon H_{\mathrm{se}}(t)]|\psi (t) \ket \mathrm{ , } \label{DinPer}
\end{equation}
where $\varepsilon$ is a small real parameter that sets the perturbation strength. In this case, from perturbation theory, the evolved state of the system is given by \cite{Sakurai:Book,Zettili:Book}
\begin{eqnarray}
|\psi (t) \ket &=& |\psi_{0} (t) \ket + \frac{\varepsilon}{i\hbar}\int_{0}^{t}U(t^{\prime})H_{\mathrm{se}}(t)|\psi_{0} (t^{\prime}) \ket dt^{\prime} \nonumber \\ &+& \left(\frac{\varepsilon}{i\hbar}\right)^{2}\int_{0}^{t}\int_{0}^{t^{\prime}}U(t^{\prime})H_{\mathrm{se}}(t)U(t^{\prime\prime})H_{\mathrm{se}}(t)|\psi_{0} (t^{\prime\prime}) \ket dt^{\prime\prime}dt^{\prime} \nonumber \\ &+& \Ocal(\varepsilon^{3})
\end{eqnarray}
where $|\psi_{0} (t) \ket$ is the ideal evolved state (unperturbed) and the $U(t)$ is the ideal propagator. For the scheme developed in this paper $U(t)=U_{1}(t)$, where $U_{1}(t)$ is given by Eq. (\ref{U1}), and $|\psi_{0} (t) \ket = U_{1}(t)| \psi _{\mathrm{inp}} \ket$ is given by Eq. (\ref{psi1Evol}). It is important to mention that in our notation we have $|\psi_{0} (\tau) \ket = U_{1}(\tau)| \psi _{\mathrm{inp}} \ket = | \psi _{\mathrm{out}} \ket$. Therefore, the probability of obtaining $| \psi _{\mathrm{out}} \ket$ can be computed from equation (up to second order) \cite{Ruschhaupt:12,X-Jing:13,Tseng:14,Kiely:14,Sakurai:Book,Zettili:Book}
\begin{equation}
P(\tau) = |\bra \psi _{\mathrm{out}} | \psi (\tau) \ket|^2 = 1 - \varepsilon^{2} \left \vert \int_{0}^{\tau} \bra \psi_{0}^{\perp} (t) |H_{\mathrm{se}}(t)| \psi_{0} (t) \ket dt \right \vert ^{2} \mathrm{ , } \label{PTau}
\end{equation}
where $| \psi_{0}^{\perp} (t) \ket$ is also a solution of the unperturbed Schrödinger equation (\ref{DinPer}) ($\varepsilon = 0$) and it satisfies $\bra \psi_{0}^{\perp} (t) | \psi_{0} (t) \ket = 0$ \cite{Ruschhaupt:12}. Through this analyzes, we can define a \textit{sensitivity} systematic error $q_{\mathrm{S}}(\tau)$ given by \cite{Ruschhaupt:12}
\begin{equation}
q_{\mathrm{S}}(\tau) = \left \vert \int_{0}^{\tau} \bra \psi_{0}^{\perp} (t) |H_{\mathrm{se}}(t)| \psi_{0} (t) \ket  dt \right \vert ^{2} \mathrm{ , }
\end{equation}
that quantifies how robust (sensitive) is the protocol against systematic errors. Therefore, a robust protocol requests a tiny value for $q_{\mathrm{S}}(\tau)$. Therefore, our aim is to minimize the function $q_{\mathrm{S}}(\tau)$ in order to maximize the fidelity $P(\tau)$ of the protocol. 

To compute $q_{\mathrm{S}}(\tau)$ associated with our protocol, we need to find $| \psi_{0} (t) \ket$ and $| \psi_{0}^{\perp} (t) \ket$. It is easy to see that the evolved state $|\psi_{0} (t) \ket$ can be obtained from Eq. (\ref{psi1Evol}) with the complex functions $\alpha(t)$ and $\beta(t)$ defined in Eq. (\ref{alpha1-beta1}), while the state $| \psi_{0}^{\perp} (t) \ket$ can be written as $| \psi_{0}^{\perp} (t) \ket = \bar{\alpha}(t)|0\ket + \bar{\beta}(t)|1\ket $, so that the condition $\bra \psi_{0}^{\perp} (t) | \psi_{0} (t) \ket = 0$ imposes that $\alpha(t)\bar{\alpha}^{\ast}(t)+\beta(t)\bar{\beta}^{\ast}(t) = 0$. In addition, once the functions $\alpha(t)$ and $\beta(t)$ depends on the input state $| \psi _{\mathrm{inp}} \ket$, it is possible to show (because the propagator $U_{1}(t)$ is unitary) that the functions $\bar{\alpha}(t)$ and $\bar{\beta}(t)$ are associated with another input state $| \psi^{\perp} _{\mathrm{inp}} \ket$, where $\bra \psi^{\perp} _{\mathrm{inp}} |\psi _{\mathrm{inp}} \ket = 0$. Thus, $\bar{\alpha}(t)$ and $\bar{\beta}(t)$ can be obtained from Eq. (\ref{alpha1-beta1}). Indeed, without loss of generality, let us write $|\psi _{\mathrm{inp}} \ket = a|0\ket + b|1\ket$, so $| \psi^{\perp} _{\mathrm{inp}} \ket = b^{\ast}|0\ket - a|1\ket$, where $a$ and $b$ are real and complex numbers, respectively. Consequently, if we change $a \rightarrow b^{\ast}$ and $b \rightarrow -a$ in Eq. (\ref{alpha1-beta1}), we can obtain $\bar{\alpha}(t)$ and $\bar{\beta}(t)$, respectively. In conclusion, in this case, we get
\begin{equation}
q_{\mathrm{S}}(\tau) = \frac{1}{4} \left \vert \int_{0}^{\tau} \omega_{x}(t) \left[ \beta(t)\bar{\alpha}^{\ast}(t)+\alpha(t)\bar{\beta}^{\ast}(t) \right] dt \right \vert ^{2} \mathrm{ . }
\end{equation}

Thus, now we can consider some particular case where we can study (analytically or not) the function $q_{\mathrm{S}}(\tau)$.

\vspace{0.3cm}

\textbf{Case one --} For simplicity of the calculations, let us consider an arbitrary single-qubit gate applied to the particular input state $| \psi _{\mathrm{inp}} \ket = | 0 \ket$. Furthermore, we are considering a gate where we have the parameters $\phi(t)= 2n\pi$, for $n \in \Z$, and $\theta(t)=\theta_{0}$ (for example, the gates discussed in Sec. \ref{SecSingleGates}).  In this case, we get
\begin{equation}
q_{\mathrm{S}}(\tau) = \frac{\sin^{2} \theta_{0}}{4} \left \vert \int_{0}^{\tau} \dot{\varphi}(t) \left[ \cos(2\theta_{0}) \sin^{2} \frac{\varphi(t)}{2} -\cos^{2} \frac{\varphi(t)}{2}  + i \cos\theta_{0} \sin\varphi(t) \right] dt \right \vert ^{2} \mathrm{ . }
\end{equation}

Firstly, it would worth to highlight that differently from others protocols of inverse engineering where the parameter $q_{\mathrm{S}}$ was studied \cite{Ruschhaupt:12,X-Jing:13,Tseng:14,Kiely:14}, in this particular case we have the parameter $q_{\mathrm{S}}$ independent on the total evolution time $\tau$. In fact, let us define the normalized time $s = t/ \tau$, so that $s \in [0;1]$. Therefore, with definition we can rewrite the above equation as
\begin{equation}
q_{\mathrm{S}} = \frac{\sin^{2} \theta_{0}}{4} \left \vert \int_{0}^{1} \frac{\varphi(s)}{ds} \left[ \cos(2\theta_{0}) \sin^{2} \frac{\varphi(s)}{2} -\cos^{2} \frac{\varphi(s)}{2}  + i \cos\theta_{0} \sin\varphi(s) \right] ds \right \vert ^{2} \mathrm{ , }
\end{equation}
therefore we have $q_{\mathrm{S}}$ independent on the total evolution time $\tau$. In addition, in this particular case we can analytically solved the above equation for any function $\varphi(t)$ as
\begin{equation}
q_{\mathrm{S}} = \frac{\sin^{2}\theta_{0}}{4} \left\{\cos^{4}\theta_{0} [ \cos \varphi(\tau) -1 ]^2 + [\cos^{2}\theta_{0}\sin \varphi(\tau) + \varphi(\tau)\sin^{2}\theta_{0}]^2 \right\} \mathrm{ . }
\end{equation}
where we have used the boundary condition $\varphi(0)=0$. Remarkably, we can see that $q_{\mathrm{S}}$ depends on the boundary conditions for parameter $\varphi(t)$. However, it is important to mention that such result is a particular result due to our consideration of the Hamiltonian $H_{\mathrm{se}}(t)$, for others Hamiltonian we can obtain a different result.

In order to show that we can optimize the protocol against the systematic error considered here, let us consider the gates discussed in this paper. Since we have considered the initial state as a computational basis state, let us discuss the sensitivity for implementing the Hadamard gate. In this case we have $\theta_{0} = \pi/4$ and $\varphi = \pi$, thus we obtain $q_{\mathrm{S}} = (8+\pi^2)/32 \approx 0.558$. Therefore, if we set $\theta(t)$ constant, the operation Hadamard gate applied to input state $|0\ket$ can be implemented with sensitivity $q_{\mathrm{S}}\approx 0.558$, since no decoherence acts.

\vspace{0.3cm}

\textbf{Case two --} Now, we will show that we can find a different value for $q_{\mathrm{S}}$ if we consider others possibilities for the parameters $\theta(t)$ and $\varphi(t)$. Thus, by considering a case where we have a time-dependent parameter $\theta(t)$, we have (by using the normalized time $s$)
\begin{equation}
q^{\prime}_{\mathrm{S}} = \frac{1}{4} \left \vert \int_{0}^{1} \sin^{2} \theta(s)\frac{\varphi(s)}{ds} \left[ \cos 2\theta(s) \sin^{2} \frac{\varphi(s)}{2} -\cos^{2} \frac{\varphi(s)}{2}  + i \cos\theta(s) \sin\varphi(s) \right] ds \right \vert ^{2} \mathrm{ . } \label{qprime}
\end{equation}

\begin{figure}[t!]
	\centering
	\subfloat[Function $P(\tau)$.]{\includegraphics[scale=0.27]{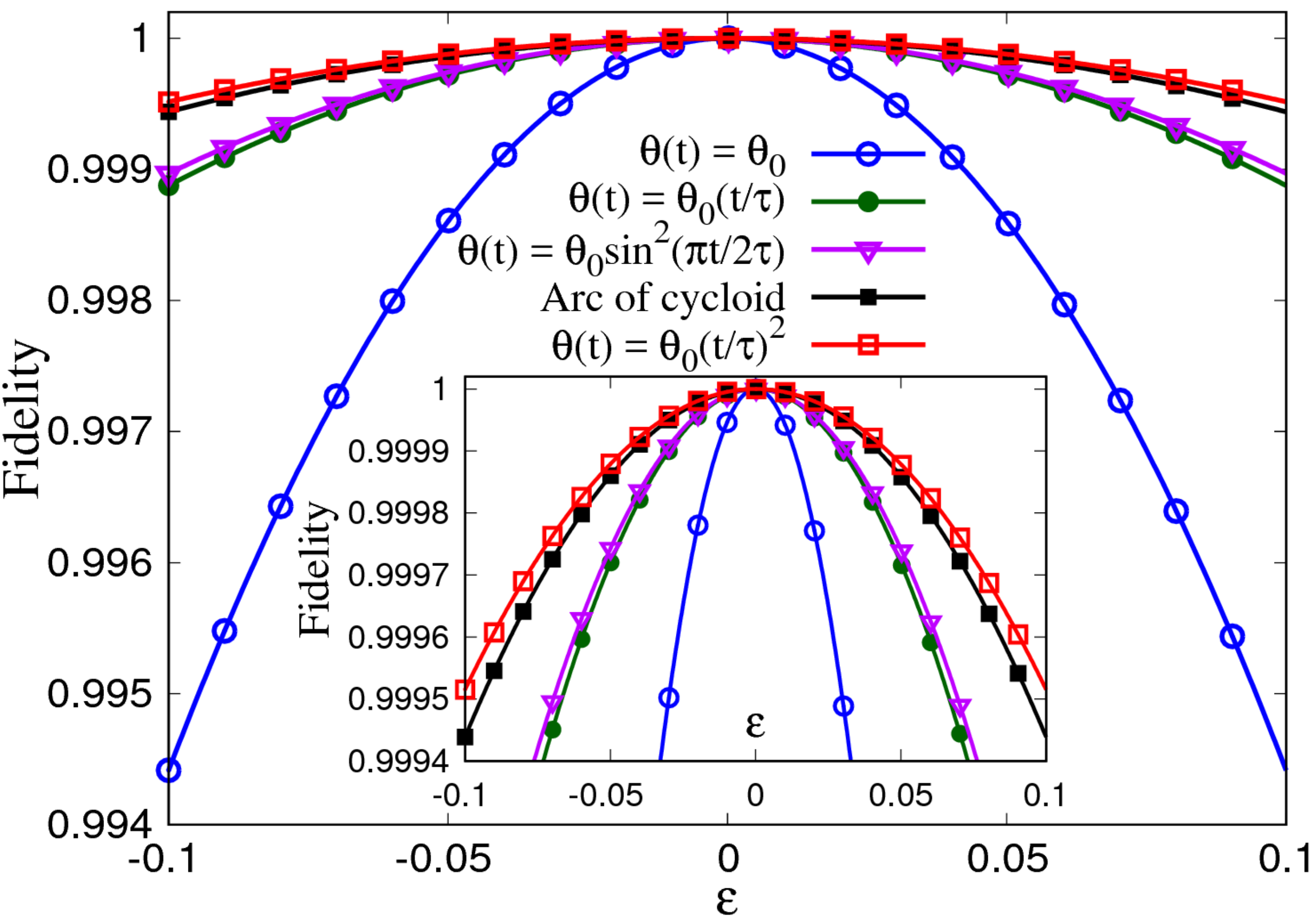} \label{Fig1}} \hspace{0.1cm}
	\subfloat[Functions $\theta(t)$, $\omega_{x}(t)$ and $\omega_{z}(t)$.]{\includegraphics[scale=0.27]{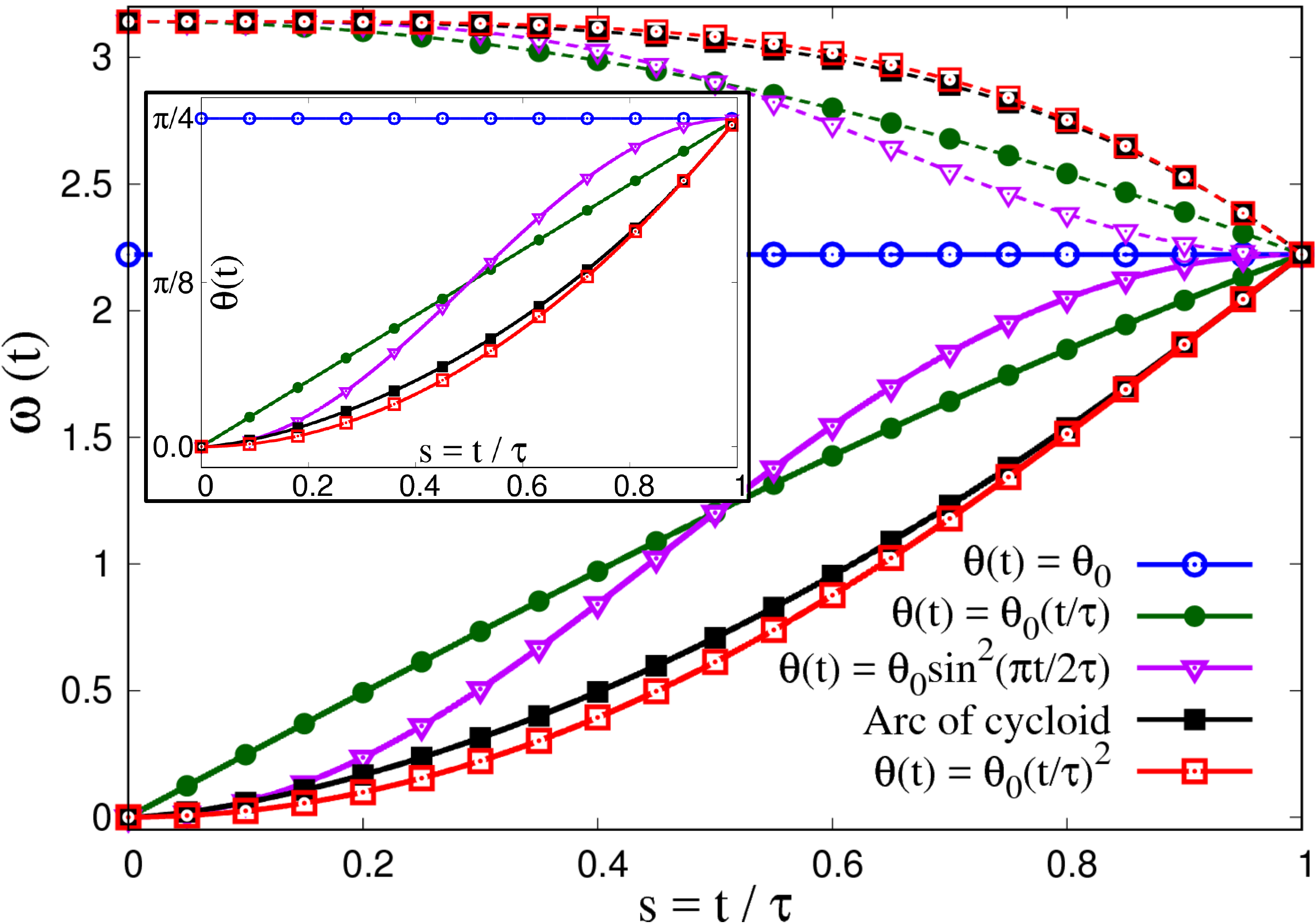}\label{Fig2}}
	\caption{(Fig. \ref{Fig1}) Fidelity $P(\tau)$ computed from Eq. (\ref{PTau}) as function of the parameter $\varepsilon$ for some choices of the function $\theta(t)$, with $\varphi(t) = \varphi_{0} t/\tau$. (Fig. \ref{Fig2} - main graph) Time-dependence of the parameters $\omega_{x}(t)$ (continuum lines) and $\omega_{z}(t)$ (dashed lines), in multiples of the total evolution time $\tau$, for some choices of $\theta(s)$. (Fig. \ref{Fig2} - inset graph). Both graphs (\ref{Fig1}) and (\ref{Fig2}) correspond to the Hadamard gate, where $\varphi_{0} = \pi$ and $\theta(\tau)=\pi/4$.}
	\label{Fig}
\end{figure}

Unlike from \textit{case one}, to solve the equation above we need to choose the functions $\theta(t)$ and $\varphi(t)$. In order to illustrate the role of the free parameters introduced here, let us keep $\varphi(t) = \varphi_{0} t/\tau$ and consider the function $\theta(t)$ as our free parameter that satisfies the boundary condition $\theta(\tau) = \theta_{0}$. Here we will consider (i) a constant function $\theta_{\mathrm{cte}}(t) = \theta_{0}$, (ii) the linear interpolation $\theta_{\mathrm{lin}}(t) = \theta_{0} t/\tau$, (iii) quadratic interpolation $\theta_{\mathrm{qua}}(t) = \theta_{0} (t/\tau)^2$, (iv) an trigonometric interpolation $\theta_{\mathrm{tri}}(t) = \theta_{0} \sin^{2}(\pi t/ 2\tau)$ and (iv) a non-trivial interpolation given by an arc of cycloid, i.e., we consider
\begin{equation}
\theta_{\mathrm{cyc}}(t) = r \arccos\left(1 - \frac{1}{r}\frac{t}{\tau}\right) - \sqrt{\frac{t}{\tau}\left(2r - \frac{t}{\tau}\right)}
\end{equation}
where $r$ is the ratio of the cycloid. However, we will use the parameter $r$ in order to satisfy the boundary condition $\theta_{\mathrm{cyc}}(\tau) = \pi/4$. In particular, we use $r \approx 0.69294$, so that $\theta_{\mathrm{cyc}}(\tau) \approx \pi/4$. The Fig. \ref{Fig2} (inset) we plot each function $\theta (t)$ considered here.

In Fig. \ref{Fig1} we show the fidelity $P(\tau)$ for each $\theta (t)$ discussed above, where we vary the parameter $\varepsilon$. We choice to vary the parameter $\varepsilon$ within the interval $[-0.1;0.1]$. Under this consideration, we take into account a systematic error so that an experimental implementation is performed with some $\omega^{\mathrm{real}}_{x}(t) \in [ \omega_{x}(t)-0.1\omega_{x}(t) \mathrm{ ; } \omega_{x}(t)+0.1\omega_{x}(t)]$. This assumption is reasonable, since the error due to imperfect calibration of the RF pulse for some experimental implementations in NMR is about $1\% \sim 10\%$ \cite{Mitra:09,Raitz:15,Bernardes:16,Isabela:16}. The Fig. \ref{Fig1} shows that the parameter $\theta(t)$ develops an interesting role for obtaining a robust protocol against errors systematic errors. Thus, as we have said, the free parameters introduced in our approach may be useful for providing robust Hamiltonians against the systematic errors considered here.

As an ``experimental guide" for understanding how the physical parameters $\omega_{x}(t)$ and $\omega_{z}(t)$ work, the Fig. \ref{Fig2} shows the behavior of the Rabi (resp. Larmor) frequency $\omega_{x}(t)$ (resp. $\omega_{z}(t)$) (always in multiples of the total evolution time $\tau$) for each $\theta(t)$ considered above. Therefore, if the total evolution time is of the order of milliseconds (microseconds), the intensity of $\omega_{x}(t)$ and $\omega_{z}(t)$ is of order of MHz (GHz). From the Figs. \ref{Fig1} and \ref{Fig2}, we can see that we can obtain a robust protocol with simple functions $\omega_{x}(t)$ and $\omega_{z}(t)$, however we can obtain an enhanced scheme with more complicated functions $\omega_{x}(t)$ and $\omega_{z}(t)$. In addition, to obtain the better function $\theta(t)$ that minimizes the sensitivity function given by Eq. (\ref{qprime}) can be a hard task.

In addition, it is important to highlight that our approach can be limited by the experimental setup used for implementing it. For instance, if we wish to implement a Hadamard gate, the protocol does not work with Hamiltonians driven by laser fields where we have the boundary conditions $\theta (0) = \theta (\tau) = 0$ \cite{Kang:16,Chen:16}. Moreover, the protocol does not work for any boundary condition where we need $\theta (\tau) \neq \pi/4$. However, as we have discussed, we can implement such protocol by using another physical systems. For example, quantum dots \cite{Fujisawa:09}, trapped ion \cite{Cui:16}, nuclear magnetic resonance \cite{Nielsen:Book} and any experimental setup where the Landau-Zener Hamiltoian can be implemented with the boundary $\theta (\tau) \neq \pi/4$.

\section{Conclusion}

In summary, in this paper we have introduced a new scheme to perform universal QC via inverse engineering of a Hamiltonian from the evolution operator. We discuss the general aspects of our approach and show how obtain a set of Hamiltonian that allow us to implement an universal set of quantum gates. Our method is an economic scheme that can be view as an alternative to others method present in the literature. In fact, while many protocols requires auxiliary qubits to perform universal QC, our approach does not need help of auxiliary elements to implement single and controlled arbitrary quantum gates. In particular, we have discussed about a restricted set of quantum gates that can be used for quantum computation. Furthermore, by using this approach, we can obtain a large class of Hamiltonians to implement single and two-quit gates where we use only two-qubit interactions.

To end, we have studied the robustness of our approach against systematic errors, due to imprecise calibration of experimental apparatus. In particular, we have considered errors related with Rabi frequency (for example, when there is deviations in the amplitude of the RF field in NMR). We show that the free parameters introduced in this paper can be useful for compensating such systematic errors, from a suitable choice of such parameters. In general the discussion considered here may not be efficient for others kind of errors, however we can obtain the ideal free parameters, for each case independently, with the intent of providing an enhanced dynamics against such errors.

\section{Acknowledgments}
	
	We acknowledge financial support from the Brazilian agencies CNPq and the Brazilian National Institute of Science and Technology for Quantum Information (INCT-IQ). We would like to thank to Natanael Moura from Universidade Regional do Cariri for help me to write this manuscript.

\section*{References}


\providecommand{\newblock}{}

\end{document}